# Utilizing Social Media Analytics to Detect Trends in Saudi Arabia's Evolving Market

**Kanwal Aalijah**

School of Electrical Engineering and Computer Science (SEECS), H-12 Campus, Islamabad, Pakistan
National University of Science and Technology, NUST Islamabad, Pakistan

**Abstract**
Saudi Arabia' faced a swift economic growth and societal transformation under Vision 2030. This offers a unique opportunity to track emerging trends in the region, which will ultimately pave the way for new business and investment possibilities. This paper explores how AI and social media analytics can identify and track trends across sectors such as construction, food and beverage, tourism, technology, and entertainment thereby helping the businesses make informed decisions. By leveraging a tailored AI-driven methodology, we analyzed millions of social media posts each month, classifying discussions and calculating scores to track the trends. The approach not only uncovered the emerging trends but also shows diminishing trends. Our methodology is able to predict the emergence and growth of trends by utilizing social media data. This approach has potential for adaptation in other regions. Ultimately, our findings highlight how ongoing, AI- powered trend analysis can enable more effective, data- informed business and development strategies in an increasingly dynamic environment.

## Introduction

The Kingdom of Saudi Arabia is experiencing a rapid change from a vision 2030 perspective and this calls for an assessment of the regional market trends at different points in time. One of the objectives of vision 2030 is to further foster economic diversification by reducing oil revenues and building new economies in construction and tourism. Moreover, the vision also expands up to the IT and communication sector, sports, entertainment and food & beverage [1]. Traditionally market surveys or interviews were used to assess the trends. The traditional methodologies are static and, therefore, fail to capture trends and shifts in consumer minds and expectations [2]. Due to the active involvement of millions of users, social networks have turned into a great source of data, that is capable of shedding light on the state of the market [3][4]. AI methods have now made it possible to perform content analysis in depth and real-time. The results from these methodologies can keep the businesses, investors, and policymakers informed of the changing dynamics in the market business environments [5].

By using Natural Language Processing (NLP) models and methodologies, vast amounts of unstructured social media data can be interpreted [5]. These methodologies can be deployed in Saudi Arabia where new socio-economic initiatives are changing some of the industries, for example, tourism, ICT, and entertainment. This type of market analysis allows public and private organizations to apply NLP and unsupervised clustering algorithms [2] to social media in a way that is able to continuously monitor and anticipate shifting marketplace dynamics. AI methodology proposed in this paper can provide sector specific analysis to stake holders and to understand the market trends at large. The analysis can help the institutions to make the right decisions. For this study data was collected from social media and NLP models such as BERT, HDBSCAN and Forecasting models were used to analyze the trends and to generate patterns. This framework can be extended to other regions as well, where the policy makers are studying the effects of new policies on the sentiment and perception of general public. This kind of research can prove to be a valuable resource for understanding and capitalizing on economic, social, and technological shifts.

## Literature Review

Traditional market research methods are now considered inadequate to cope with rapid changes in the dynamics of modern market. They are lagging behind the bilateral dynamics and volatility in the markets; especially in the case of Saudi Arabia [2]. Traditional market research methodologies fail as they are unable to keep up with changes in consumer thoughts or sentiments in the high growth regions which are undergoing drastic socio-economic changes. With the help of AI, trends can now be analyzed efficiently in real- time. This can be done through the data collected from social media networks [4]. The data collected from social media can be fed in to machine learning models which can identify patterns by clustering the data and insights can be generated [6]. Advances in NLP especially in LLMs has now made it possible to understand the changing market and consumer dynamics [7]. Since the region is transforming hence consumer behaviors bound to change, hence it is essential for Saudi Arabia to monitor trends. Most of the data in KSA is generated in Arabic language. Applying models such as BERT [8] and AraBERT [9] has been shown to analyze modern Arabic unstructured text with greater accuracy [10].

AI-based analytics are being used for trend analysis focusing on evolving customer habits and shifts in various sectors and industries[11]. Just like other countries, even in Saudi Arabia, social media serves as the base line platform for public discourse. It serves as a tool to provide insights into sector- specific trends such as in tourism, construction, and food & beverage industry [12]. Data generated through social media platforms can be fed in to machine learning models to generate insights and predictions. Models like Latent Dirichlet allocation and BERTopic are generally use to understand the context and organize the unstructured data [2]. Deeper understanding of consumer sentiment can be obtained by using new and advance deep learning models such multimodal sentiment analysis [13]. These models are able to incorporate text and visual content both. This will ultimately lead to a deeper understanding of consumer sentiment. It will be very beneficial for the decision makers, if they have the ability to continuously monitor and analyze trends over time.

There is a dire need of an AI methodology, that can bridge the gap made by traditional research methods. This methodology should be robust enough to track the sentiment, emotions of people towards an initiative or a problem under consideration. In this paper we propose a machine learning frame work that takes advantages of various NLP models to generate insights, track sentiment thereby helping the business and policy makers to better understand the evolving KSA socio economic market.

## Methodology
### A. Data Collection
Data was sourced from major social media platforms, including X, TikTok, Instagram, and Facebook. These clustering these embeddings using a class-based TF-IDF (c- TF-IDF) [17] approach. Once the topics were identified we clustered the common topics together using HDBSCAN (Hierarchical Density-Based Spatial Clustering of Applications with Noise) [18] for clustering, as it does not require pre-specifying the number of clusters, allowing the data to naturally reveal topic groupings.

Once the overall topics in the entire dataset were identified, we assigned topics to each post.

After that we segmented the dataset into three time periods—2014-2017, 2018-2020, and 2021-202.

D. Trend Popularity and growth trajectory metrics

In order to understand perception of public towards a trend, we consolidated the engagement (likes, shares, and comments) and sentiment metrics and proposed a formula that tracks the current popularity of a trend. By using this metric we were able to measure a trend's relevance and potential for growth in upcoming years [11].

platforms were selected for their popularity and high engagement in Saudi Arabia. This enabled us to capture real- time public sentiment and sector-specific discussions. We collected over 30 million posts, comments, and videos from

where:

$$PPt = \sum N\ 1\ Si\ X\ E \quad (1)$$

2014 to 2024. The posts that had geo location set as KSA or they had certain hashtags like #saudi etc. were considered relevant and others were discarded. The engagement data such as likes, saves, shares, comments etc. was also maintained.

Data preprocessing involved several steps: removing noise (URLs, emojis), normalizing text (standardizing

Arabic script), tokenizing text, filtering stop words, and applying stemming and lemmatization to handle Arabic morphology. Additionally, n-grams (bigrams and trigrams) were generated to capture context and common phrases within the short social media posts.

## B. Sentiment Anaysis

Social media data in Saudi market is mostly generated in Arabic language. Moreover, Saudi Arabic has various dialects which makes sentiment analysis difficult. These dialects, along with informal language and slang, often deviate from Modern Standard Arabic, adding complexity to language processing tasks. To address this, we used AraBERT [9], a BERT [8] based models. AraBERT was used as it specialized for Arabic, BERT was used to handle other languages. The reason for using BERT is it effectively handles dialectal variations and colloquial expressions commonly found in social media [10], and has been bench marked against many traditional models and methodologies [9].

AraBERT Model was fine-tuned on a dataset of regional Arabic language with training split into 80% for training, 10% for validation, and 10% for testing. Cross-entropy loss [14] and the Adam optimizer [15] were used during fine-tuning to improve model performance while preventing overfitting.

## C. Topic modeling and clustering

In our research, we utilized BERTopic [16],which is a cutting-edge topic modeling technique that employs BERT embeddings [16] to analyze topics from a dataset. It captures the semantic nuances of the text, providing a sophisticated understanding of underlying topics. The process involves transforming the dataset into BERT embeddings, followed by $Si$ is the sentiment score for each post i

$Ei$ is the engagement level (sum of likes, shares,comments)

$N$ is the total number of posts in the given month

## E. Predict trend growth

To predict the growth of a trend for upcoming years, we used Prophet Model [19]. It is a simple model, that handles seasonality and trend shift with great accuracy. Moreover, the model has been benchmarked against ARIMA, LSTM and it was proven that Prophet model handles seasonality well [19]. Hence it was our ideal choice, given the nature of the study.

Results

In this section we are going to present the results that are obtained by following above methodology.

Data collection and pre-processing

Our dataset included over 30 million social media posts from X, Instagram, and Facebook, collected between 2014 and 2024 as show in Table I. These platforms were chosen for their widespread use and engagement in Saudi Arabia. After various pre-processing steps, the collected data was divided in to year buckets. Reason of these buckets was to understand the shift of trends before the launch of Vision 2030 and after the launch of Vision 2030.

Table I. : Collected data

| Time Range | No. of posts |
| --- | --- |
| 2016-2017 | 3 million |
| 2018-2020 | 9 million |
| 2021-2024 | 17 million |

## Sentiment Analysis

To analyze sentiment in our dataset, we first separated the posts into Arabic and non-Arabic texts. For the Arabic posts, we used AraBERT. For non-Arabic posts, we applied a standard BERT model that supports multiple languages, including English. Both models classify the sentiment of each post as positive, neutral, or negative. Fig 1 shows the sentiment data was then visualized to show the proportion of positive, neutral, and negative sentiments, weighted by the number of posts.

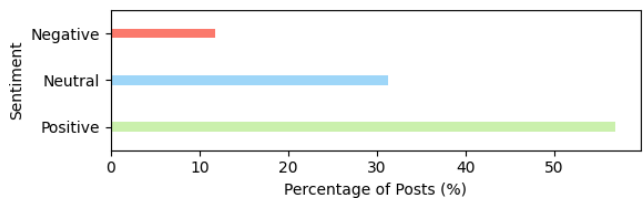

Fig. 1. Sentiment Distribution of the posts

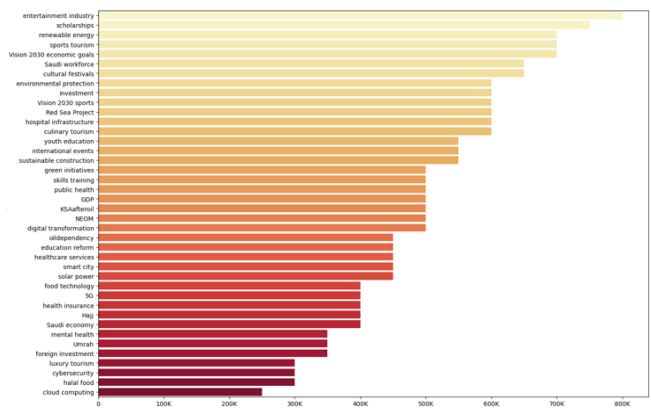

Fig. 2. Number of posts collected for each time range

**Identification and clustering of Trends**
The BERTopic model uncovered a variety of topics in the dataset. The initial topics can be seen in the Fig. 2. It only a subset of the topics that were initially identified.

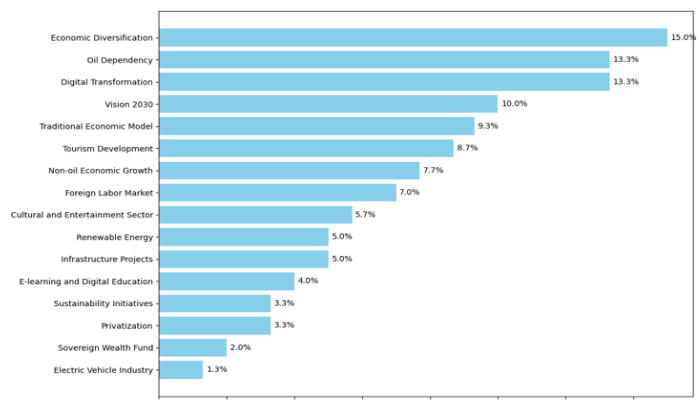

Fig. 3.  A snapshot of identified topics

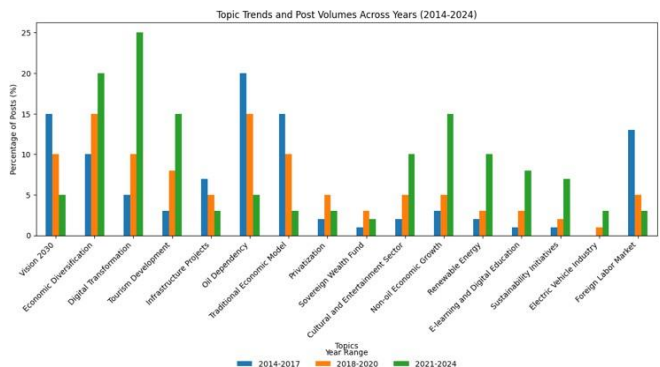

Fig. 4. A snapshot of identified topics per year

The topics were then clustered to determine the optimal number of topics that best represent the dataset. Fig 3 shows the final topics. Some of the irrelevant topics were discarded along with their posts as well. For instance, the final topic 'Sustainability Initiatives" emerged from discussions on renewable energy, electric vehicle infrastructure etc. This clustering method enabled the simplification of complex community discussions into clear, actionable topics.

Each post was assigned a topic. In certain cases, more than one topic was also assigned to a post. Fig 4 shows, the topics diversification by time bucket. After assigning topics to the posts, their engagement levels were studies along with the sentiment.

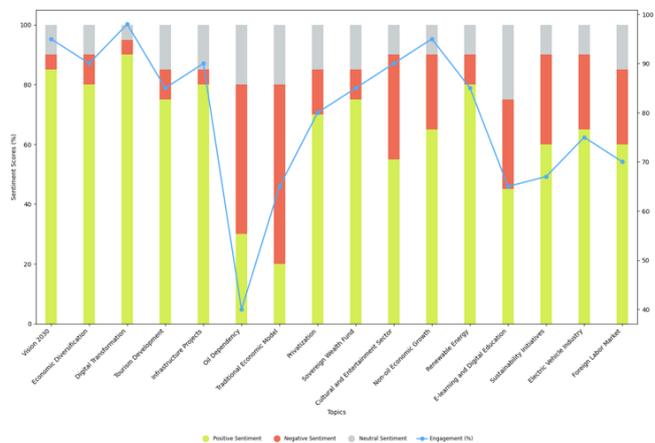

Fig. 5. Sentiment vs Engagment vs Topics

Fig 5 shows a drop in engagement trends for oil dependency topic. The percentage of negative sentiment for this topic is also quite high. This shows clearly, as per Vision 2030, KSA has started to move away from Oil dependency and we can see it is also evident from the collected data, hence that shows public is also steering in the similar direction along with the leaders. Digital transformation show notable high engagement, it was also seen in Fig 4 that as compared with other topics, Digital Transformation grew in the 2021-2024 period and the engagement shows the similar pattern. Traditional Economic Model also showed a significant decrease in the more recent years, also indicating the shift away from oil in Saudi Arabia's development strategy.

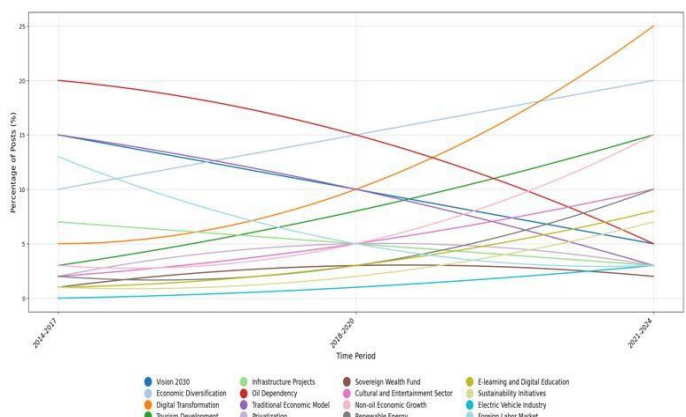

Fig. 6. Trend Popularity

## Trend Popularity

In order to predict the emerging trends, we used the data from the historical periods we first calculated Trend popularity values using the equation (1). Fig 5 shows the calculated trend popularity for each topic and plotting it against time.

The trend popularity values obtained, were then used to predict the growth of trend in upcoming year. We used Prophet forecasting model to handle the prediction. The model calculates the relationship between the years and the post volumes, engagement etc. then predicts the value for upcoming years onwards based on that trend. This model uses the post volumes in the previous periods to predict the post volume for the next

period (2024-2027).

Fig 6 shows the Emerging Trends predicted. The dots actually shows where the trend will reach in the upcoming year. It can be observed in the figure, some trends are dropping and some are growing. For example, Topics like Vision 2030, Digital Transformation, and Economic Diversification exhibit consistent growth. Furthermore, there is a sustained growth in the popularity of concepts such as Sustainability Initiatives and Renewable Energy, which is expected stay positive in the near future. In contrast, the topics of the reduction in oil dependency and Traditional economic model are likely to shift towards a diminishing trend. The reason for this could be as a part of Vision 2030, it was decided that KSA will move away from these topics. The trends in the dataset provide an insight in the changing priorities of the public towards the new age concepts of modernization and sustainability.

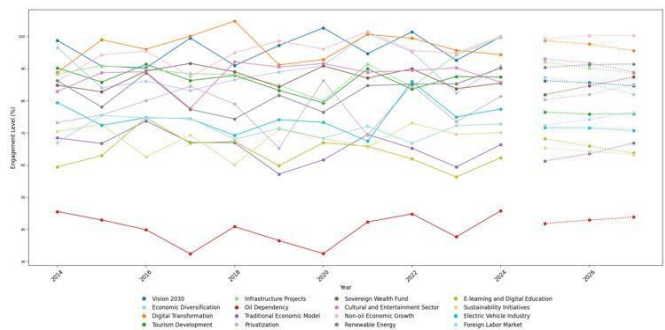

Fig. 7. Trend Growth Prediction

In contrast, the articles of the Drying oil dependency and Traditional economic model are likely to shift towards a diminishing trend for the reasons that these are increasingly irrelevant to the dataset. The trends in the dataset provide an insight in the changing priorities of the public towards the new age concepts of modernization and sustainability.

**Conclusion and contribution**

This paper presents an innovative methodology for forecasting trend, combining sentiment analysis and engagement metrics from social media data. Key contributions include enhanced trend forecasting through the integration of sentiment scores and engagement data, providing businesses, investors, and policymakers with a reliable measure of trend relevance and growth potential. The paper introduces a dynamic trend growth prediction model that not only tracks current trends but also forecasts future growth by considering momentum and the rate of change, offering more accurate predictions than traditional models.

The process can be updated in real-time by setting up cron jobs, that can continuously update the Pulse Potential scores allow for continuous tracking and forecasting, enabling businesses and investors to stay ahead in fast-changing markets like Saudi Arabia. Moreover, the model not only helps in identifying the emerging trends but can also discover the diminishing trends in the given data. It offers critical insights for investment, new businesses and marketing strategies. This methodology can be applied to other areas as well such as urban planning, economic policy development etc. thereby helping the decision makers by providing real- time feedback on public sentiment to make informed decisions.

**Limitations and future work**

Although the research is able to generate promising results but there are several limitations. The model's results are heavily dependent on the availability and quality of social media data. The results will always show what is in the collected dataset. Our process is only able to collect a subset of dataset from these platforms. There is a possibility for the results to be biased. Moreover, we are tracking limited social media platforms and offline events are not being tracked in this study. The offline events can also influence public opinion.

This study provides valuable insights into trend forecasting through the Pulse Potential (PP) metric, but there are several areas for future research. First, exploring micro- trends within macro trends, such as breaking down digital transformation into subcategories like AI, cloud computing, and cybersecurity, could provide a more granular view of trend evolution. Additionally, sentiment-driven forecasting could be enhanced by segmenting stakeholders (e.g., consumers, businesses, government) to better capture their distinct perceptions of trends, such as how economic diversification is viewed by different groups. Cross-platform data integration, incorporating data from news articles, forums, and blogs, could offer a more holistic view of trends, while dynamic weighting of engagement metrics would optimize the predictive power of likes, shares, and comments based on trend context. Model's accuracy can be further increased, if external factors like economic shits, political changes, global crisis etc. are incorporated. Moreover, if geo- data is also incorporated in the dataset, we could further perform analysis based on location-specific predictions. This will be beneficial in identifying the trends that may be more prominent in urban areas like Riyadh as compared with the rural regions.

**Disclaimer**

This study is conducted solely for research purpose. This study is not affiliated with or does not intend to promote any business or commercial activity. The only objective of this study is to explore how AI, can be used to generate insights into market trends by using social media data. The findings presented in this paper are based on the data available on social media platforms during the study period.